%% file: paper.tex
\documentclass[conference]{IEEEtran}

\IEEEoverridecommandlockouts

\usepackage[show]{imports/chato-notes}

\usepackage{cite}
\usepackage{amsmath,amssymb,amsfonts}

\usepackage{graphicx}
\usepackage{textcomp}
\usepackage{xcolor}

\usepackage{algorithm}
\usepackage{algpseudocode}
\usepackage{booktabs}
\usepackage{float}
\usepackage{multirow}
\usepackage{bm}
\usepackage{hyperref}

\usepackage{url}
\usepackage{listings}
\usepackage{enumitem}

\lstdefinelanguage{Rust}{
    sensitive=true,
    morekeywords=[1]{break,continue,else,for,if,in,loop,match,return,while,as,box,const,enum,extern,fn,impl,let,mod,mut,pub,ref,self,static,struct,trait,type,unsafe,use,where,crate,false,true},
    morekeywords=[2]{i8,i16,i32,i64,i128,isize,u8,u16,u32,u64,u128,usize,f32,f64,str,char,bool,String,Vec,Option,Result,Self},
    morecomment=[l]{//},
    morecomment=[s]{/*}{*/},
    morestring=[b]",
    moredelim=[is][\color[RGB]{150, 80, 0}]{'}{'}, 
}

\lstdefinestyle{ruststyle}{
    language=Rust,
    basicstyle=\ttfamily\scriptsize,
    keywordstyle=\color[RGB]{185, 0, 0}\bfseries,
    keywordstyle=[2]\color[RGB]{30, 100, 150},
    stringstyle=\color[RGB]{0, 120, 0},
    commentstyle=\color[RGB]{120, 120, 120}\itshape,
    emph={'a,'b,'v,'e,'q},
    emphstyle=\color[RGB]{150, 80, 0},
    numbers=left,
    numberstyle=\tiny\color{gray},
    stepnumber=1,
    numbersep=5pt,
    showstringspaces=false,
    breaklines=true,
    frame=single,
    backgroundcolor=\color[RGB]{250, 250, 250},
    lineskip=-1.5pt,
    aboveskip=5pt,
    belowskip=5pt,
    xleftmargin=12pt
}

\usepackage{amsmath}

\DeclareMathOperator*{\argmin}{arg\,min}
\input{sections/10-macros}

\input{imports/revision_commands}

\def\BibTeX{{\rm B\kern-.05em{\sc i\kern-.025em b}\kern-.08em
    T\kern-.1667em\lower.7ex\hbox{E}\kern-.125emX}}
    
\begin{document}

\title{Compact and Efficient Indexes\\for Learned Sparse Retrieval
}

\author{
\IEEEauthorblockN{Franco Maria Nardini}
\IEEEauthorblockA{\textit{ISTI--CNR}\\
Pisa, Italy\\
francomaria.nardini@isti.cnr.it}
\and
\IEEEauthorblockN{Luca Rizzo}
\IEEEauthorblockA{\textit{University of Pisa} \\
Pisa, Italy\\
l.rizzo11@studenti.unipi.it}
\and
\IEEEauthorblockN{Cosimo Rulli}
\IEEEauthorblockA{\textit{Linkup\IEEEauthorrefmark{1}\thanks{\IEEEauthorrefmark{1}Work done while at ISTI--CNR, Pisa, Italy.}}\\
Paris, France\\
cosimo@linkup.so}
\and
\IEEEauthorblockN{Rossano Venturini}
\IEEEauthorblockA{\textit{University of Pisa}\\
Pisa, Italy\\
rossano.venturini@unipi.it}
}

\maketitle

\input{sections/00-abstract}
\input{sections/01-introduction}
\input{sections/02-relatedwork}

\input{sections/03-methodology}

\input{sections/04-experiments}

\input{sections/05-conclusions}

\newpage
\section*{Acknowledgments}
The authors were partially supported by the European Union's Horizon Europe Research and Innovation program, under the RenAI project (GA No. 101291378) and by the ENRESFOOD project, funded under the FutureFoodS partnership. The project is co-funded by the Austrian Science Fund (FWF; 10.55776/KIN4661525), The Dutch Ministry of Agriculture, Fisheries, Food Security and Nature (BO-43-219-026), the Italian Ministry of University and Research (MUR), the German Federal Ministry of Research, Technology and Space (031B1717), and the European Union's Horizon Europe research and innovation programme via FutureFoodS, the European Partnership for a Sustainable Future of Food Systems (GA No. 101136361).

In accordance with IEEE guidelines on the use of generative AI tools, we disclose that generative AI technologies, i.e., Grammarly and ChatGPT 5.5 Instant, were used to assist in grammar checking and language polishing during the preparation of this paper. All research ideas, experiments, analyses, and writing were conducted and critically reviewed by the authors. No part of the scientific content or creative reasoning has been generated or substantially rewritten by generative AI tools.

\bibliographystyle{IEEEtran}
\bibliography{biblio}
\end{document}

%% file: sections/10-macros.tex
\newcommand{\msmarco}{\textsc{Ms Marco}\xspace}

\newcommand{\seismic}{\textsc{Seismic}\xspace}

\newcommand{\hnsw}{\textsc{Hnsw}\xspace}

\newcommand{\kannolo}{\textsc{kANNolo}\xspace}
\usepackage{adjustbox}

\newcommand{\sindi}{\textsc{Sindi}\xspace}

\newcommand{\supblock}{\textsc{Dsp}\xspace}

\newcommand{\splade}{\textsc{Splade}\xspace}

\newcommand{\spladeT}{\textsc{Splade}-v3\xspace}
\newcommand{\epic}{\textsc{Epic}\xspace}

\newcommand{\etal}{\emph{et al.}\xspace}

\newcommand{\oursshort}[1]{\sc{Pfi}}

\newcommand{\infless}{\textsc{LiLsr}\xspace}

\newcommand{\fma}{\texttt{fma}\xspace}

\newcommand{\hnswsvbdp}{\textsc{Hnsw} [dotp] }
\newcommand{\qdrant}{\textsc{Qdrant}\xspace}

\newcommand{\codec}[1]{\textsc{#1}}
\newcommand{\gammaelias}{\codec{Elias-$\gamma$}\xspace}
\newcommand{\deltaelias}{\codec{Elias-$\delta$}\xspace}
\newcommand{\rice}{\codec{Rice}\xspace}
\newcommand{\expgolomb}{\codec{Exponential-Golomb}\xspace}
\newcommand{\zetaen}{\codec{Zeta}\xspace}
\newcommand{\vbyte}{\codec{VByte}\xspace}
\newcommand{\for}{\codec{For}\xspace}
\newcommand{\qmx}{\codec{Qmx}\xspace}

\newcommand{\simdbp}{\codec{Simd-Bp128}\xspace}
\newcommand{\pfor}{\codec{PFor}\xspace}

\newcommand{\optpfor}{\codec{OptPFor}\xspace}

\newcommand{\simplesixteen}{\codec{Simple16}\xspace}
\newcommand{\simple}{\codec{Simple}\xspace}
\newcommand{\simpleeight}{\codec{Simple8b}\xspace}
\newcommand{\bp}{\codec{Bp32}\xspace}
\newcommand{\eliasfano}{\codec{Elias-Fano}\xspace}
\newcommand{\clusteredeliasfano}{\codec{Clustered Elias-Fano}\xspace}
\newcommand{\bic}{\codec{Binary Interpolative Coding}\xspace}
\newcommand{\dotpacking}{\codec{DotPacking8}\xspace}
\newcommand{\dotpackingblocked}{\codec{DotPacking8-J}\xspace}
\newcommand{\dotpackingdp}{\codec{DotPackingDp}\xspace}
\newcommand{\cdotpacking}{\codec{CDotPacking8}\xspace}
\newcommand{\cdotpackingdp}{\codec{CDotPackingDp}\xspace}
\newcommand{\cexpgolomb}{\codec{CExp-Gol}\xspace}
\newcommand{\dotvbyte}{\codec{DotVbyte}\xspace}
\newcommand{\parteliasfano}{\codec{Partitioned Elias-Fano}\xspace}
\newcommand{\uncompressed}{\codec{Plain-u16}\xspace}
\newcommand{\jumpdot}{\textsc{JumpDot}\xspace}



%% file: imports/revision_commands.tex
\usepackage{soul}
\usepackage{xcolor}

%% file: sections/00-abstract.tex
\begin{abstract}
This paper investigates how to substantially reduce the memory footprint
of learned sparse retrieval indexes without sacrificing the efficiency of
state-of-the-art retrieval data structures. Building on \seismic, we
revisit both levels of its design: the inverted index used to select
candidates and the forward index used to score them. For the inverted
index, we replace costly per-block summaries with \emph{medoids}, namely
existing documents elected as block representatives, collapsing the
per-block metadata from a sparse vector to a single document identifier.
For the forward index, we compress both components and values. We reorder the vocabulary to place co-occurring components closer together and encode the resulting $\Delta$-gaps with \dotpacking, a SIMD-friendly bit-packing scheme that fuses decompression with dot-product evaluation; values are quantized with compact per-component 4-bit codebooks fitted to each component's distribution. We further introduce \jumpdot, a blocked dot-product kernel tailored for queries that contain only a few non-zero entries. Our forward-index compression is independent of \seismic{} and can be plugged into any system relying on forward-index-based scoring, as we demonstrate by integrating it into \kannolo. A comprehensive evaluation on
\msmarco{} with three state-of-the-art learned sparse encoders shows that our solutions markedly improve the speed--space trade-off of learned
sparse retrieval: at equal accuracy, our indexes answer queries up to
$5.3\times$ faster than the best competitor while using about $3\times$ less memory, and in the most memory-constrained regime, they remain up to $1.9\times$ faster while using up to $3.9\times$ less memory.

\end{abstract}

\begin{IEEEkeywords}
learned sparse retrieval, compression, efficiency.
\end{IEEEkeywords}

%% file: sections/01-introduction.tex
\section{Introduction}
\label{sec:introduction}

Vector search is a core primitive in modern data-management and machine-learning systems, where applications from semantic search to retrieval-augmented generation reduce to finding the vectors most similar to a query in a learned space. For text, \emph{learned sparse representations} (LSR) are especially attractive: they encode queries and documents in a high-dimensional space but activate only a few dimensions, each with a non-zero weight~\cite{splade-sigir2021,formal2021splade,lassance2024spladev3,epic,formal2023tois-splade,yates2024neural}. LSR attains very high effectiveness~\cite{formal2022splade,lassance2024spladev3,formal2023tois-splade,10.1007/978-3-031-28241-6_7,nguyen2026sparton} while staying interpretable---active dimensions are vocabulary terms with contextualized importance weights---and generalizing well out of domain through learned query and document \emph{expansion}, which mitigates the classic \emph{vocabulary mismatch} problem~\cite{thakur2021beir,formal2022splade,lassance2024spladev3}.

Despite their effectiveness, LSR vectors are hard to index: their activated dimensions and weights depart from the skewed Zipfian patterns that classical inverted indexes exploit~\cite{mackenzie2021wacky,bruch2023sinnamon,bruch2023bridging}, so traditional machinery struggles to deliver low-latency top-$k$ retrieval. This was reflected in the 2023 Big ANN Sparse Track~\cite{simhadri2024resultsbigannneurips23}, whose winning solutions were adaptations of the graph-based \textsc{Hnsw}~\cite{hnsw2020} rather than inverted-index engines. Since 2024, a wave of methods tailored to LSR---\seismic~\cite{bruch2024seismic,Bruch2024SeismicWave,ecir-seismic,bruch2025efficientsketchingnearestneighbor}, \textsc{Bmp}~\cite{mallia2024faster}, \supblock~\cite{carlson2025dynamic}, and \sindi~\cite{li2025sindi}---has cut retrieval latency from hundreds of milliseconds to hundreds of microseconds while keeping search almost exact. For instance, \seismic answers queries on \splade-encoded \msmarco in about $200\,\mu$s single-threaded, yet requires a data structure of about $7$ GiB. As latency enters the microsecond regime, memory rather than latency becomes the first-order bottleneck, echoing the history of inverted-index compression~\cite{pibiri2020techniques}, making the footprint of these structures the natural next frontier.

Modern LSR indexes typically follow a two-level design. A \emph{selector index} rapidly identifies a small set of promising candidate documents, while a \emph{forward index} stores the sparse document representations and supports the dot products needed to score the selected candidates. In \seismic, for example, the selector consists of pruned inverted lists equipped with block summaries, whereas the forward index stores the sparse vectors used to compute exact scores. Making learned sparse retrieval compact, therefore, requires reducing the memory footprint of both components.


In this paper, we study how to make learned sparse retrieval indexes substantially more compact while preserving the efficiency achieved by state-of-the-art data structures. We build on \seismic and revisit its selector-index/forward-index components. On the selector side, we propose a memory-efficient alternative to the costly block summaries used by \seismic. In detail, we replace the summaries with \emph{medoids}, i.e., existing documents selected as representatives of the corresponding blocks. This reduces the per-block metadata from a sparse vector to a single document identifier. 
We also revisit the dot-product computation used during candidate reranking. While densified scoring~\cite{bruch2024seismic} is effective for expanded-query encoders, it becomes less suitable for inference-free retrievers, where queries have a small support (i.e., contain only a few non-zero entries), and documents, conversely, are highly expanded. We introduce a blocked dot-product algorithm (\jumpdot) that combines support-proportional computation with SIMD-friendly block matching, enabling query-driven skipping of non-matching document blocks. 
On the forward-index side, we build on the compression framework of Bruch \emph{et al.}~\cite{bruch2026forward} and further reduce the space needed to store document vectors by compressing both their components and their values. For components, we exploit the fact that document supports are stored as sorted sequences of identifiers: after reordering the vocabulary to make co-occurring components closer, we encode the resulting $\Delta$-gaps with SIMD-friendly bit-packing schemes tailored to the short and locally variable sequences produced by learned sparse encoders. For values, we fit quantizers to the value distribution of each component; we study both uniform scalar quantization and centroid-based non-uniform quantization, with a particular focus on compact 4-bit representations. The resulting design targets the main sources of memory consumption in modern learned sparse retrieval indexes while retaining the ability to efficiently identify and score candidates.  
Since the forward index is not specific to \seismic, our compression is orthogonal to the selector index and can be plugged into other forward-index-based systems; we demonstrate this by integrating it into \kannolo~\cite{kannolo}, an \hnsw implementation supporting sparse vectors, which is the second-fastest solution in our experiments.

We conduct an extensive experimental evaluation on
\msmarco~\cite{nguyen2016msmarco} with three learned sparse encoders, \splade~\cite{formal2021splade}, \infless~\cite{nardini2025effective}, and \spladeT~\cite{lassance2024spladev3}, covering different sparsity regimes and query/document length values. We compare against state-of-the-art learned sparse retrieval indexes, including \sindi~\cite{li2025sindi}, \kannolo~\cite{kannolo}, and the original \seismic, and evaluate latency, retrieval effectiveness, and memory consumption under multiple memory budgets.

The results show that our compressed indexes substantially improve the
speed--space trade-off. At equal accuracy, \seismic is up to $5.3\times$ faster than \sindi{} while using about
$3\times$ less memory; in the most memory-constrained regime, it remains
up to $1.9\times$ faster while using up to $3.9\times$ less memory.
Moreover, applying the same forward-index compression to \kannolo{}
nearly halves its memory footprint on \spladeT and reduces it by
41\% on \splade, without any deterioration in efficiency.

To summarize, the novel  contributions of this paper are:
\begin{itemize}[leftmargin=*]
    \item We propose a \emph{medoid}-based selector index for \seismic. Instead of storing a synthetic per-block summary vector, we select an existing document of the block as its representative, collapsing the per-block metadata to a single document identifier, thus shrinking the inverted index.
    
    \item We introduce \jumpdot, a blocked dot-product algorithm for regimes where only queries are extremely sparse, like inference-free retrievers, where queries contain only a few non-zero entries while documents remain highly expanded. \jumpdot combines support-proportional computation with SIMD-friendly block matching, enabling query-driven skipping of non-matching document blocks. 

    \item We design a compressed forward index for learned sparse vectors. We compress component identifiers by reordering the vocabulary and encoding the resulting $\Delta$-gaps with fine-grained bit-packing extended to support SIMD-integrated decoding and scoring over the short, locally variable supports of LSR vectors, yielding \dotpacking and \dotpackingdp, their reference-compressed variants \cdotpacking and \cdotpackingdp, and \dotpackingblocked, which integrates \jumpdot with compressed-block skipping. We compress values with per-component quantization, studying both uniform scalar and centroid-based $4$-bit codebooks. Being independent of the selector, this contribution can be plugged into any forward-index-based retrieval engine.

    \item  We conduct an extensive experimental evaluation on \msmarco with three learned sparse encoders---\splade, \infless, and \spladeT---covering different sparsity regimes, and against state-of-the-art baselines (\sindi, \kannolo, and \seismic). The results show that our techniques improve the speed--space trade-off beyond \seismic: \seismic preserves practical query efficiency while moving to substantially lower memory regimes, and the same forward-index compression makes \kannolo markedly more compact while retaining its latency profile without sacrificing efficiency. On \infless, the \jumpdot kernel and its compressed variant \dotpackingblocked further makes \seismic the fastest and most memory-efficient specialized method.

\end{itemize}

%% file: sections/02-relatedwork.tex

\section{Related Work}
\label{sec:related}

Learned sparse representations (LSR) encode queries and documents as high-dimensional, sparse vectors whose active dimensions correspond to vocabulary terms~\cite{zamani2018sparse,sparterm,zhao-etal-2021-sparta,gao2021coil,unicoil}. The \splade family~\cite{splade-sigir2021,formal2021splade,formal2022splade,lassance2024spladev3}, together with models such as \epic~\cite{epic} and efficiency-oriented variants~\cite{lassance2022efficient-splade}, has established LSR as a state-of-the-art paradigm for first-stage text retrieval, combining strong effectiveness with the interpretability and out-of-domain robustness afforded by learned term expansion~\cite{formal2023tois-splade}. A more recent line further lowers query-time cost through \emph{inference-free} retrievers, which replace the query encoder with a learned token-impact lookup table, so that no neural inference is run at query time~\cite{nardini2025effective,10.1145/3726302.3730192}. Across these models, a common obstacle remains: LSR encoders produce term distributions that depart markedly from the Zipfian patterns of natural text, which makes them difficult to serve efficiently with classical inverted-index machinery~\cite{mackenzie2021wacky,bruch2023sinnamon,bruch2023bridging}.

\vspace{1mm}
\noindent\textbf{\seismic and its descendants}.
\seismic~\cite{bruch2024seismic,Bruch2024SeismicWave,ecir-seismic,bruch2025efficientsketchingnearestneighbor} is the state-of-the-art retrieval machinery for LSR. It exploits the \emph{concentration of importance} exhibited by learned sparse encoders---only a few coordinates of each vector dominate the inner product---to organize every inverted list into geometrically cohesive \emph{blocks} obtained through inner-product clustering, and equips each block with a \emph{summary} vector that conservatively upper-bounds the dot product attainable by the documents it references. At query time, the summary acts as a selector: a block is fully evaluated via a forward index storing the document vectors only when its summary score survives a dynamic-pruning test. \seismic has since inspired several systems that refine the candidate-selection stage. Block-Max Pruning (\textsc{Bmp})~\cite{mallia2024faster} organizes the index into blocks of documents equipped with max-impact bounds; Dynamic Superblock Pruning (\supblock)~\cite{carlson2025dynamic} groups blocks into superblocks to prune at a coarser granularity; and \sindi~\cite{li2025sindi} is an approximate maximum-inner-product index that scans impact-sorted lists with query and document-side pruning. Across these designs, the selector relies on auxiliary per-block metadata---summaries or impact bounds---whose storage grows with the number of blocks and constitutes a sizable fraction of the index; \sindi, moreover, stores component identifiers and values uncompressed, which places it firmly in the high-memory regime. Our work targets exactly this overhead: rather than storing a synthetic summary vector per block, we elect an existing document, the block \emph{medoid}, as its representative, collapsing the per-block metadata to a single identifier.

\vspace{1mm}
\noindent\textbf{Toward compact indexes}.
As retrieval latency for LSR has entered the microsecond regime, memory has become the primary bottleneck. This mirrors the trajectory of classical inverted indexes, where achieving practical query times was followed by a sustained effort to reduce their space footprint~\cite{pibiri2020techniques}. The forward index of learned sparse systems, however, is still commonly stored either uncompressed or with a single, collection-wide scalar quantizer. Bruch \etal~\cite{bruch2026forward} recently introduced \dotvbyte, a SIMD-integrated codec that compresses forward-index components while supporting fast dot-product evaluation, establishing that compression and scoring need not be decoupled. We build on this insight along two axes. For components, we combine vocabulary reordering via recursive graph bisection~\cite{dhulipala2016compressing,mackenzie2021faster} with bit-level, SIMD-decodable bit-packing codecs~\cite{pibiri2020techniques} tailored to the short sequences of LSR vectors, fusing decompression and scoring within the vector units. For values, we move from collection-level to per-component quantization, fitting the quantizer to each term's value distribution. Because the forward index is shared by several LSR systems, this part of our contribution can be plugged into other forward-index-based retrieval engines.

\vspace{1mm}
\noindent\textbf{Approximate nearest neighbor search}.
A complementary body of work casts sparse retrieval as approximate maximum inner product search and adapts data structures originally devised for dense vectors. Graph-based indexes such as \textsc{Hnsw}~\cite{hnsw2020}, its inner-product variant \textsc{Ip-Nsw}~\cite{ip-nsw18}, and \textsc{DiskAnn}~\cite{diskann} build navigable proximity graphs over the collection; the two winning solutions of the 2023 BigANN Sparse Track~\cite{simhadri2024resultsbigannneurips23} were in fact adaptations of \textsc{Hnsw} to sparse vectors. We rely on \kannolo~\cite{kannolo}, an efficient \textsc{Hnsw} implementation supporting both dense and sparse domains, to build the graph-based baseline used in our evaluation. A second line partitions the collection into clusters and scans only those closest to the query, building on cluster pruning~\cite{chierichetti2007clusterPruning} and cluster-skipping inverted indexes~\cite{Altingovde2008ClusterSkipping,Cinar2023ClusterSkipping}; \textsc{Sinnamon}~\cite{bruch2023sinnamon} applies this idea, together with a coordinate sketch, to streaming sparse vectors.

\vspace{1mm}
\noindent\textbf{Vector quantization and sketching}.
To reduce memory usage and accelerate distance computation, dense approximate search has long relied on lossy vector compression: product quantization~\cite{pq} and its optimized~\cite{opq}, locally optimized~\cite{locallyOptimizedPQ}, anisotropic~\cite{scann}, and multiscale~\cite{multiscaleQuantization} variants, which compress entire vectors into compact codes. These techniques quantize a vector jointly across its dimensions; in contrast, we quantize each coordinate independently with a per-component codebook, exploiting the small vocabulary of LSR encoders to fit a dedicated quantizer to every term's value distribution. Closer to sparse retrieval, dimensionality reduction and sketching reduce the number of non-zero coordinates while approximately preserving inner products: random-projection and sampling-based sketches for inner-product estimation~\cite{woodruff2014sketching,daliri2023sampling}, the coordinate sketch of \textsc{Sinnamon}~\cite{bruch2023sinnamon}, and the Set $\alpha$-MSS sketch used by \seismic~\cite{bruch2025efficientsketchingnearestneighbor} all reduce the effective dimensionality of sparse vectors prior to scoring.

%% file: sections/03-methodology.tex

\section{Methodology}
\label{sec:methodology}

We consider a dataset $\mathcal{X}$ of sparse vectors $x_i \in \mathcal{X}$.
Each $x_i$ is composed of a set of components $c_i \in C $ and a set of real---non-negative---values $v_i \in V$ weighting the importance of $c_i$ in $x_i$.
We define $s_i$ as the support of $x_i$, namely the components in $x_i$ with $v_i \neq 0$.
With this in mind, the vector $x_i$ can be written as a set of pairs $x_i = \{ (c^p_i, v^p_i), p \in s_i \}$.
In the following, when the index $i$ is not needed in the discussion, we will refer to $x_i$ as $x$ to ease notation. 
We also consider $V_{c} = \{ v_j^{(c)} | \ j = 0, \ldots, |V_C| \}$ as all the possible values associated with a component $c \in C$ in $\mathcal{X}$.
As done for $x_i$, when the index $i$ is not needed in the discussion, we will refer to $c_i$ as $c$.
Moreover, we also define $q_i$ as the sparse query vector, which is given to find its top-$k$ nearest neighbors in $\mathcal{X}$ according to a given similarity function, e.g., cosine or dot product.


\subsection{\seismic}
\label{subsec:seismic}

\seismic~\cite{bruch2024seismic,Bruch2024SeismicWave,ecir-seismic,bruch2025efficientsketchingnearestneighbor} is the state-of-the-art machinery for retrieval over learned sparse representations.
Following the two-level design outlined in Section~\ref{sec:introduction}, it combines a \emph{selector index}, which rapidly identifies a small set of promising candidate documents for the query, with a \emph{forward index}, which stores the sparse vectors $x_i \in \mathcal{X}$ and supports the exact dot-product computations needed to score those candidates. We describe the two components in turn.

\vspace{1mm}
\noindent\textbf{Selector index}.
The selector is a pruned and blocked inverted index. For each component $c \in C$, \seismic first builds the inverted list $\mathcal{I}_c$ of the identifiers of all vectors $x_i$ with $c \in s_i$, sorted in decreasing order of value $v$. It then applies a three-step procedure, each step governed by a hyper-parameter:
\begin{enumerate}[leftmargin=*]
\item each $\mathcal{I}_c$ is truncated to $\lambda$ entries on average, discarding the postings that contribute least to the inner product;
\item each pruned list is partitioned into at most $\beta$ geometrically-coherent \emph{blocks} through a shallow, inner-product-based variant of $k$-means: $\beta$ centroids $\{\mu^{(1)}, \dots, \mu^{(\beta)}\}$ are sampled uniformly at random from $\mathcal{I}_c$, and each $x \in \mathcal{I}_c$ is assigned to the block $B_{c, j^\ast}$ whose centroid maximizes the dot product with it, $j^\ast = \arg\max_{1\leq r \leq \beta}\, \langle x, \mu^{(r)} \rangle$;
\item each block $B$ is equipped with a \emph{summary} vector $\phi(B)$, defined component-wise as the largest value attained inside the block, $\phi(B)[c] = \max_{x \in B,\; c \in s(x)} x[c]$ (and zero otherwise).
\end{enumerate}

Since both query and data vectors are non-negative, $\langle q, \phi(B) \rangle \geq \langle q, x \rangle$ for every $x \in B$: the summary score is a conservative \emph{upper bound} on the inner product attainable inside the block. A block can therefore be safely skipped whenever $\langle q, \phi(B) \rangle$ falls below the current top-$k$ threshold.
The support of $\phi(B)$ is the union of its members' supports, which would inflate the index; \seismic thus (i) replaces $\phi(B)$ with its \emph{$\alpha$-mass subvector} $\tilde\phi(B)$, keeping only the largest entries whose cumulative value covers a fraction $\alpha \in (0,1]$ of $\|\phi(B)\|_1$, and (ii) scalar-quantizes the surviving values to a single byte by mapping $[m_B, M_B]$ to $256$ equal-width bins, where $m_B$ and $M_B$ are the per-block minimum and maximum.

\vspace{1mm}
\noindent\textbf{Forward index}.
The forward index stores the complete sparse vectors of the collection. When the selector decides that a block must be visited, \seismic retrieves the exact representation of each of its documents from the forward index and computes the true inner product with the query, updating the top-$k$ heap accordingly. 

\vspace{1mm}
\noindent\textbf{Memory footprint}.
Making \seismic compact requires shrinking \emph{both} parts. The forward index grows with the total number of non-zero entries in the collection. The selector, in turn, adds one summary vector per block on top of the inverted lists: even though each summary is sketched and quantized, their number scales with the total number of blocks, and their cumulative size is far from negligible.
On the \splade-on-\msmarco index discussed in Section~\ref{sec:introduction}---whose data structure occupies about $7$ GiB---the roughly $5$M block summaries, each retaining around $100$ non-zero components ($\sim$$400$ bytes), account for nearly half of the index.
This observation drives our two contributions: we compress the selector by replacing summaries with \emph{medoids} (Section~\ref{subsec:medoids}), and we compress the forward index along both its components and its values (Sections~\ref{subsec:compcompr}~and~\ref{subsec:valcompr}). The former is motivated by a natural question: do we really need a synthetic per-block summary $\phi(B)$ at all?

\subsection{\seismic with medoids}
\label{subsec:medoids}
We answer the previous question negatively by replacing the summary vector built by \seismic with an \emph{existing} document from the block itself as its representative. By doing so, the per-block summary collapses to a single document identifier, whose coordinates are read on demand from the forward index, where they are already stored. This approach is based on the hypothesis that, for sufficiently small and coherent blocks, a single document is sufficient to discriminate relevant blocks from irrelevant ones.
As a consequence, the memory footprint shrinks from $\sim$400 bytes to a single $8$-byte document identifier. This implies that, from a memory consumption perspective, with the same space requirements, we can now afford a finer blocking, i.e., more and smaller blocks per list, without paying the price of the increased storage that original \seismic summaries would require. Since blocks are built by using inner-product $k$-means, a natural choice is to pick as the representative of $B$ the member that is itself closest to the block centroid under the same similarity, i.e.\ the document $\mathrm{med}(B) = \arg\max_{x \in B} \langle x, \bar{B} \rangle$, where $$\bar{B} = \tfrac{1}{|B|} \sum_{x \in B} x$$ is the component-wise mean of the block. We refer to $\mathrm{med}(B)$ as the \emph{medoid} of $B$.

Unlike $\phi(B)$, the medoid does not allow to define an upper bound on the dot product that can be achieved between the incoming query and the documents of the block. For any query $q$, $\langle q, \mathrm{med}(B)\rangle$ sits inside the spread of $\{\langle q, x\rangle : x \in B\}$, and it is therefore an \emph{estimation} of the block-max score $\max_{x \in B}\langle q, x\rangle$---possibly arbitrarily smaller---rather than the conservative overestimate that $\phi(B)$ provides. Blocks whose true maximum lies well above their medoid score can therefore be wrongly skipped. To recover recall if this case happens, we exploit the same heap-factor mechanism already exploited by \seismic, which requires entering any block whose medoid score exceeds $\texttt{heap\_factor}\cdot\theta_k$, where $\theta_k$ is the running top-$k$ threshold and $\texttt{heap\_factor}\in(0,1)$ is a bouncing factor that compensates for the looser pruning signal.

Similarly to what happens in \seismic, we begin scanning each posting list by computing the dot product between the query and all the medoids in that list. As medoid vectors are stored in the forward index, we rely on standard dot product computation rather than on column-major SIMD scan over the summary matrix as done by the original \seismic. This leads to several implications. First, we lose the ability to exploit SIMD batching. Second, medoids are scattered in memory, and every dot product computation incurs random forward-index accesses and L3 cache misses. A solution to the two problems mentioned requires the use of prefetching, which means we batch $8$ medoids at a time, we issue all prefetches up front, and we then compute the $8$ dot products, so that the memory accesses run in parallel and are overlapped with the computation on the same batch.

\subsection{Efficient Dot Product Computation}
\label{subsec:dotprod}

Since sparse vectors store only their non-zero entries, the dot product $\langle q, x \rangle = \sum_{p \in s_q \cap s_x} q^p \cdot x^p$ reduces to a weighted intersection of the two supports. We consider different ways to implement it. 

\vspace{1mm}
\noindent \textbf{Merge-based dot product}.
Assume the components of $q$ and $x$ are stored in strictly ascending order. The two $(c^p_q, v^p_q)$ and $(c^p_x, v^p_x)$ of the query and the document, respectively, are traversed with a two-pointer merge: at each step we advance the pointer with the smallest component index, and whenever the two indices coincide we accumulate the product $q^p \cdot x^p$ via an \fma into a scalar accumulator. The procedure runs in        
$\mathcal{O}(|s_q| + |s_x|)$ time, touches only the non-zero entries of both operands, and has a memory footprint proportional to the supports rather than to $|C|$. Its main drawback is the poor predictability of its control-flow, which limits the effectiveness of vectorization and exposes the algorithm to branch-mispredictions.

\vspace{1mm}
\noindent \textbf{Densified dot product}.
This approach treats the sparse query as a dense vector. We
allocate an array $d_q \in \mathbb{R}^{|C|}$, initialized to zero, and scatter the query values into it: $d_q[c^p] \leftarrow v^p$ for every $p \in s_q$. The dot product with any document $x$ is then   
$\langle q, x \rangle = \sum_{p \in s_x} d_q[c^p_x] \cdot v^p_x$, i.e., a single linear scan over the document support in which each component index $c^p_x$ is used as a direct lookup into $d_q$.
The inner loop becomes branch-free, vectorizes naturally (the lookups map to gather instructions), and runs in $|s_x|$ per document, plus the densification of the query, amortized as it happens once per search.
In terms of memory used, storing $d_q$ grows with $|C|$. This is acceptable for moderate vocabularies, e.g., the $\sim$30k WordPiece tokens of \splade, but starts to significantly degrade when $|C|$ exceeds the L3 cache, since lookups become essentially random accesses to main memory.

\vspace{1mm}
\noindent \textbf{\jumpdot}.
The two dot product computation methods introduced, i.e., merge-based and densified methods, sit at opposite ends of a trade-off: while the former is compact but induces branches in the computation, and it is hard to vectorize, the latter vectorizes naturally but pays an $|C|$-size memory footprint that collapses once $d_q$ no longer fits in cache.
\jumpdot yields the best of both approaches. It retains the support-proportional footprint of the merge while recovering the vectorization of the dense scan. It works by storing the document $x$ as a sequence of fixed-size blocks of $L$ consecutive non-zero entries. To do so, we use $L=16$ so that the $L$ component indices, stored as $16$-bit integers, and the $L$ half-precision values of a block require exactly one $64$-byte cache line (unused slots in the last block are padded with zeroes).
The dot product is then a two-pointer merge between the query terms and the document \emph{blocks}: we scan the blocks of $x$ in order as we advance through the sorted query terms. Using the per-block maximum component, we discard a block in a single comparison if its largest index is below the current query component, stopping early once the query is exhausted. For each block that may contain a match, we broadcast the current query component and compare it against all $L$ component indices of the block with one SIMD instruction; the (at most one) matching value is selected by masking, converted to single precision, and accumulated through an \fma into a vector accumulator, which is horizontally reduced once all blocks have been processed. This collapses control flow to roughly one branch per block instead of one per entry, and resolves each query term against $B$ document entries with a single $B$-wide SIMD comparison, for a cost of $\approx |s_q| + |s_x|/L)$ vector operations. Like the merge, it touches only the non-zero entries and uses memory proportional to $|s_q| + |s_x|$ rather than $|C|$, so it does not degrade for large vocabularies; unlike the merge, its predictable block-granular control flow and SIMD matching make it especially effective for short queries over longer documents, where block-skipping discards most of the document at negligible cost.

\subsection{Forward Index Components Compression}
\label{subsec:compcompr}
Since sparse embedding vectors store only non-zero entries, the component identifiers form a strictly increasing sequence. Our objective is to compress these sequences to achieve a performance profile comparable to the state-of-the-art \dotvbyte~\cite{bruch2026forward} in dot-product evaluation speed, while significantly reducing the average number of bits per component. The analysis initially focuses on identifying the most efficient compressors in terms of bits per integer. Subsequently, it is crucial to evaluate the impact on end-to-end query latency by integrating these compression techniques within the \seismic framework.

We first compress monotonic component sequences using $\Delta$-gaps encoding. To identify the most promising encoders, we treat all $\Delta$-gaps as if generated by a unified zero-order source, matching empirical frequencies to the encoders' ideal distributions. By modeling embeddings and components as a bipartite graph, we further optimize this distribution via Recursive Graph Bisection (RGB)~\cite{dhulipala2016compressing}, which partitions components to yield a permutation that places co-occurring terms close to each other. Using the Mackenzie \etal~\cite{mackenzie2021faster} implementation, this reordering drops the empirical Shannon entropy from 8.29 to 6.92, the median gap ($p_{50}$) from 72 to 13, and the $p_{75}$ from 202 to 63. Since a median gap of 13 requires only 4 bits, the byte-aligned \dotvbyte (allocating 9 bits per element up to 255) introduces a 5-bit overhead for half the components, motivating transitioning to finer-granularity bit-level or bit-packing schemes.

The reason above leads us to design a comprehensive analysis of different encoders that follows the hierarchical taxonomy by Pibiri and Venturini~\cite{pibiri2020techniques}, which is organized around three primary levels of compression.
In the first level, we find \textit{integer codes} encoding each component independently, including universal (\gammaelias, \deltaelias~\cite{elias1975universal}), parametric (\rice~\cite{rice1971some}, \expgolomb, \zetaen~\cite{boldi2005zeta}), and byte-aligned (\vbyte) schemes.
The second category comprises list-level compressors, which can be further divided into two main approaches: block-based encoders and increasing sequence-level encoders. The former encodes blocks of components independently, as seen in \bp~\cite{lemire2015decoding}, the \simple family~\cite{zhang2008simple16, anh2010inverted}, and various \codec{PForDelta} variants~\cite{zukowski2006super, yan2009inverted}. The latter exploits the monotonic nature of the data, and includes techniques such as \eliasfano~\cite{elias1974efficient, fano1971number}, \parteliasfano~\cite{ottaviano2014partitioned}, and \bic~\cite{moffat2000binary}.
The third level encompasses \textit{index compressors} that leverage collection-wide redundancies, such as the clustered reference-based approach~\cite{pibiri2017clustered}, adapted for forward index components.
Within the second category, SIMD-based bit-packing encoders like \simdbp~\cite{lemire2015decoding} and \qmx~\cite{trotman2014qmx} are of particular interest. These methods are especially relevant as they leverage SIMD instructions to enable low-latency dot product evaluation, potentially matching the efficiency of \dotvbyte while delivering superior compression density due to a bit-packing approach.

It is important to note that while traditional integer compression schemes are optimized for long inverted lists, the document embeddings in LSR models are relatively short: in the \msmarco \spladeT encoding, $p_{50}$, $p_{75}$, and $p_{95}$ are 166, 204, and 270, respectively.

This fine-grained nature introduces a new challenge for existing sequence encoders. Methods like \optpfor partition sequences into blocks of 128 integers, assuming statistical regularity within that window. Similarly, SIMD-based encoders such as \simdbp and \qmx utilize large blocks to amortize metadata and maximize throughput. However, since the median length of document embeddings is only 166 in the \spladeT encoding of \msmarco, these techniques often group almost the entire sequence into a single block, failing to capture the local homogeneity of the data and leading to bit-waste. Consequently, we require encoders that balance density with high-speed scoring by adapting to the local characteristics of short sequences.

\vspace{1mm}
\noindent\textbf{Space compression evaluation}.
We conducted space simulations using high-performance libraries (\texttt{fastpfor-rs}~\cite{fastpforrs}, \texttt{dsi-bitstream}~\cite{dsibitstream}, and \texttt{ds2i}~\cite{ds2i}) to identify the most efficient compression strategies. 
The first BPI column of Table~\ref{tab:space_results} reports results for reordered component gaps, whose pooled distribution has empirical Shannon entropy 6.92 bits per gap. As shown there, parametric scalar codecs like \expgolomb adapt well to the skewed remapped distribution, approaching the entropy lower bound. Among block-based schemes, \simplesixteen outperforms \simpleeight and \bp by adjusting its bit-packing configuration more frequently,  isolating outliers that would otherwise force larger blocks into inefficient bit-widths. 

While \optpfor utilizes an exception-handling mechanism, its reliance on 128-integer blocks proves too coarse for the high local variability of our dataset. Conversely, block-based SIMD encoders like \simdbp and \qmx exhibit the poorest density. \simdbp operates on large blocks of 128 integers, a granularity too coarse for document vectors averaging only 170 components: a single outlier forces the entire block into a wider bit-width. Similarly, \qmx fails to match \simplesixteen because its low-bitwidth selectors require large sequences (e.g., 40 integers for $b=3$) to saturate the registers, a regularity that short document bursts cannot provide.

\begin{table}[t]
    \caption{Bits per integer on \spladeT--\msmarco for reordered component gaps and clustered reference-based compression.}
    \label{tab:space_results}
    \centering
    \footnotesize
    \setlength{\tabcolsep}{4pt}
    \begin{tabular}{lrr}
        \toprule
        \textbf{Codec} & \textbf{BPI} & \textbf{Clustered BPI} \\
        \midrule
        \expgolomb & 7.25 & 6.22 \\
        \zetaen & 7.45 & 6.69 \\
        \dotpackingdp & 7.50 & 6.73 \\
        \dotpacking & 8.24 & 7.35 \\
        \simplesixteen & 8.35 & 7.68 \\
        \deltaelias & 8.42 & 7.74 \\
        \gammaelias & 8.72 & 7.86 \\
        \simpleeight & 8.82 & 8.00 \\
        \optpfor & 8.94 & 9.27 \\
        \rice & 9.02 & 6.53 \\
        \vbyte & 9.32 & 9.26 \\
        \qmx & 9.77 & 8.53 \\
        \bp & 9.88 & 9.43 \\
        \dotvbyte & 9.96 & 10.10 \\
        \pfor & 11.35 & 9.68 \\
        \simdbp & 12.25 & 9.85 \\
        \bottomrule
    \end{tabular}
\end{table}

On strictly increasing sequences of remapped component identifiers of the \spladeT encoding of \msmarco, \bic achieves an excellent result of 7.06 bits per integer by exploiting dense local clusters via recursive partitioning, whereas \eliasfano achieves 9.30 bits per integer, failing to leverage proximity due to its fixed-bit allocation.

However, achieving high density is only half the problem. Utilizing these encoders as standalone decoders follows a \textit{decompress-then-process} paradigm: they reconstruct the entire sequence into an intermediate memory buffer before the evaluator can compute the dot product. As our evaluations will show, even state-of-the-art SIMD decompressors fail to match the performance of \dotvbyte. This is because \dotvbyte is not merely a decoder but a \textit{SIMD-integrated scorer}. The decoupling in traditional encoders introduces redundant memory traffic, preventing the system from overcoming the memory wall and maximizing its throughput during search.

\vspace{1mm}
\noindent\textbf{SIMD-integrated scorer pipeline}.
Inspired by \dotvbyte, the \textit{integrated gather-driven pipeline} adapts the previously described densified dot product approach to the SIMD execution model. The key component is the AVX2 \texttt{\_mm256\_i32gather\_ps} intrinsic: passing an \texttt{\_\_m256i} register of 8 decompressed identifiers retrieves 8 weights from the densified query array. These weights are then multiplied by the corresponding embedding values and accumulated via a Fused Multiply-Add (FMA) operation. Finally, a horizontal reduction on the accumulator yields the dot product score.

\vspace{1mm}
\noindent\textbf{\dotpacking}.
To extend the SIMD-integrated decoding and scoring pipeline of \dotvbyte to fine-grained bit-packing, we propose \dotpacking, an encoder inspired by \bp. It is optimized to compress 8-element blocks and rapidly decompress them into a 256-bit SIMD register to match the specifications of the 8-lane gather.
Unlike \simdbp, which adopts a vertical layout by bit-packing contiguous elements across independent streams to enable simple intra-lane \texttt{vpsrld}, \texttt{vpand}, and \texttt{vpor} instructions to decode the integers, \dotpacking utilizes a horizontal layout.
In this horizontal layout, $\Delta$-gaps are stored consecutively in the memory bitstream. We store the 4-bit selectors (encoding bit-width $b$) separately from the payload. Since each eight-element block occupies $8b$ bits, block starts are byte-aligned, and a single 128-bit load contains the entire block for $b \le 16$. Each loaded payload is broadcast to a 256-bit register and decoded via a deterministic \textit{Spread-Shuffle-Shift-Mask} pipeline.
The SIMD operations rely on specific masks retrieved from a precomputed table based on the block selector, resulting in a fully branchless execution flow.

\vspace{1mm}
\noindent\textbf{\dotpackingblocked}.
While the 8-integer blocks produced by \dotpacking are natively suited for the gather-driven densified dot product, the layout also integrates seamlessly with the \jumpdot approach described in Section~\ref{subsec:dotprod}, which is inspired by and adapted from the V1 SIMD intersection algorithm proposed by Lemire \emph{et al.}~\cite{lemire2016simd}. Specifically, this blocked evaluation mode offers significant performance benefits for embedding models that typically generate sparse queries containing very few non-zero components. To support efficient block skipping, the layout of \dotpackingblocked stores the maximum component of each 8-integer block uncompressed in a header at the beginning of each compressed representation. During query processing, the evaluator evaluates the query components one at a time, comparing each component with the precomputed block maxima. If the query component exceeds the block's maximum component count, the entire block is skipped. In our horizontal layout, this skip is executed rapidly
so as to completely avoid the decompression of the block components. If the query component is instead less than or equal to the maximum, the block is decompressed and evaluated in SIMD registers.

\smallskip
\noindent\textbf{Dynamic partitioning with \dotpackingdp}.
While fixed-width blocks ensure regularity, local variability can lead to bit waste: a single large outlier forces the entire block to use a wider bit-width. To bridge this gap, we propose \dotpackingdp, which treats vector partitioning as an optimization problem solved via Dynamic Programming (DP). Each block uses an 8-bit selector for bit-width $b$ and length $l \le 16$. In \dotpackingdp, variable lengths $l$ lead to bit-unaligned starts; since SIMD loads are byte-addressable, subsequent blocks may re-read bytes from previous payloads, destroying the assumption that a single 128-bit load always captures the next 8 elements. To sustain high-throughput scoring, we impose a hardware boundary constraint during DP: each segment of up to 8 integers must fit entirely within a 128-bit loading window ($s + \min(l, 8) \cdot b \le 128$, where $s$ is the bit-offset). For $l > 8$, this condition must hold for both the initial 8-element block and the subsequent residual. This ensures that each selector yields either a single register (for $l \le 8$) or a pair (for $l > 8$), both guaranteed to be perfectly contained within at most two deterministic SIMD loads. The globally optimal partitioning is identified by the recurrence:
\begin{equation}
	C[i, s'] = \min_{\substack{1 \le l \le 16 \\ s \in \{0, \dots, 7\} \\ \texttt{safe}(s, b_{i,l}, l)}} \Big\{ C[i-l, s] + 8 + (l \cdot b_{i,l}) \Big\}
\end{equation}
where $b_{i,l}$ is the max bit-width in $[i-l+1, i]$, and $s'$ is the resulting offset.
During scoring, an overflow-buffered zipping mechanism merges decoded chunks into eight-element SIMD batches, retaining unused elements for the next batch.

On the sparse embedding components of the \spladeT encoding of \msmarco after remapping and $\Delta$-gap encoding, 
\dotpackingdp achieves a tighter compression ratio of 7.50 bits per integer than \dotpacking, which achieves 8.24 bits per integer. This shows that \dotpackingdp closely approaches the entropy lower bound and confirms that a finer granularity outweighs the metadata overhead.

\smallskip
\noindent\textbf{Clustered reference-based compression}.
To exploit structural redundancies across different embeddings, we employ a \textit{Clustered Reference-Based} approach inspired by \clusteredeliasfano~\cite{pibiri2017clustered}. 
The collection is partitioned using bisecting k-means with IDF-weighted component supports to group embeddings with shared components, giving less weight to globally frequent ones. For each cluster, a \textit{reference list} $R$ captures its most frequent components.
Each embedding is differentially encoded by splitting its components into a \textit{mapped} part, consisting of the subset of its components that appear in the reference list $R$ associated with the cluster it belongs to, and a \textit{residual} part for all remaining components. This effectively performs a dimensionality reduction for the mapped components since they are represented by their positions within $R$, reducing the coding universe from $|C|$ to $m = |R|$. While a larger $m$ increases coverage, the resulting expansion of the local universe toward the global vocabulary size $|C|$ gradually erodes the space savings per element.
Crucially, for each individual embedding, this process yields two distinct, strictly increasing sequences that can be compressed independently using any integer encoder.

While the original framework~\cite{pibiri2017clustered} employs \parteliasfano to compress both mapped and residual portions of each embedding and relies on \texttt{next\_geq} for intersection with the query, this approach is suboptimal for short document vectors. In a regime characterized by short sequences, the overhead of descending into the fine-grained, bit-level mechanics of \texttt{next\_geq} becomes counterproductive. While such skipping structures are highly effective for exceptionally long lists where they can bypass massive blocks of unneeded data, in this low-cardinality regime, a bulk decompression mechanism that sweeps through the entire sequence is significantly faster than executing branch-heavy, element-by-element random accesses.

The second BPI column of Table~\ref{tab:space_results} reports, for each encoder, the weighted average cost of independently compressing the mapped and residual sequences with that encoder. We fix the reference-list size to 512 components per cluster for all methods. The weighted average of the Shannon entropies of their pooled $\Delta$-gap distributions is 5.85 bits per gap, with weights proportional to component counts.
While partitioning significantly reduces the weighted entropy to 5.84, it introduces a double indirection bottleneck during evaluation: mapped components represent local positions in $R$ rather than global identifiers. To access a global component, the system must first decode the mapped position, use it to index the uncompressed reference list, and then retrieve the corresponding weight from the densified query array using the global identifier.
To extend the SIMD-integrated decoding and scoring pipeline to clustered reference-based compression, we introduce \cdotpacking and \cdotpackingdp, which bypass scalar reference lookups through a \textit{double-gather} SIMD strategy. A first integer gather on the reference list R retrieves global identifiers; a second float gather then uses them to fetch query weights for FMA multiplication and accumulation.
This keeps the entire mapped evaluation within SIMD registers, bridging reference-based space efficiency and hardware-accelerated scoring.

\subsection{Forward Index Values Compression}
\label{subsec:valcompr}
Prior work on forward-index compression~\cite{bruch2026forward} quantizes values with a single, distribution-agnostic scalar quantizer applied uniformly to the whole collection. Per-component quantization is substantially more accurate: each component $c$ has its own non-zero value distribution, so a quantizer fitted to the per-component value set $V_c$ avoids the resolution loss that a single global range incurs. This is affordable because the vocabulary of sparse encoders is small---\splade, for instance, inherits BERT's WordPiece tokenizer with $\approx30{,}000$ tokens---so the per-component metadata (a few scalars, or a short codebook, per $c$) is negligible compared to the forward index itself.

\smallskip
\noindent\textbf{Scalar Quantization}.
Scalar quantization represents a value of component $c$ as
$
v \approx \bar{v} \;=\; m_c + w_c \, k,
$
where $m_c = \min V_c$, $M_c = \max V_c$, the step size is
$w_c = (M_c - m_c)/(2^{\text{nbits}}-1)$, and the stored integer
code is $k = \operatorname{round}\!\bigl((v-m_c)/w_c\bigr)
\in \{0,\ldots, 2^{\text{nbits}}-1\}$.
Let $\bar{x}$ denote the vector $x$ with its values quantized this way.
The dot product with a query $q$ becomes
\begin{align*}
\langle q, \bar{x}\rangle
\;=\; \sum_{c \in s_q \cap s_x} q^c\,\bar{v}^c
\;=\; \sum_{c \in s_q \cap s_x} q^c m^c \;+\; q^c w^c k^c .
\end{align*}
Na\"ively, each matched component now costs three multiplications and one addition, against the single multiplication of an \texttt{f16} or \texttt{f32} representation. However, $q^c m^c$ and $q^c w^c$ depend only on the query and can be precomputed once and stored in two $|C|$-sized arrays---the same densified layout used in
Section~\ref{subsec:dotprod}---which comfortably fit in the lower cache levels. Each component then contributes a multiply-add, but the two operations cannot be fused into a single \fma because the outer \fma is already consumed by the cross-component accumulation.
The ReLU upstream of the \splade-style encoder enforces $v \geq 0$, hence $m_c \geq 0$ for every component. Empirically, on \splade and \spladeT the average $m_c$ is below $10^{-4}$, so we drop the offset entirely and quantize as $\bar{v} = w_c \, k$. The inner loop then collapses to a single \fma per matched component, which is the same cost as \texttt{f16} or \texttt{f32}, modulo the cost of widening codes to floats.

\begin{figure}
    \centering
        \includegraphics[width=\linewidth]{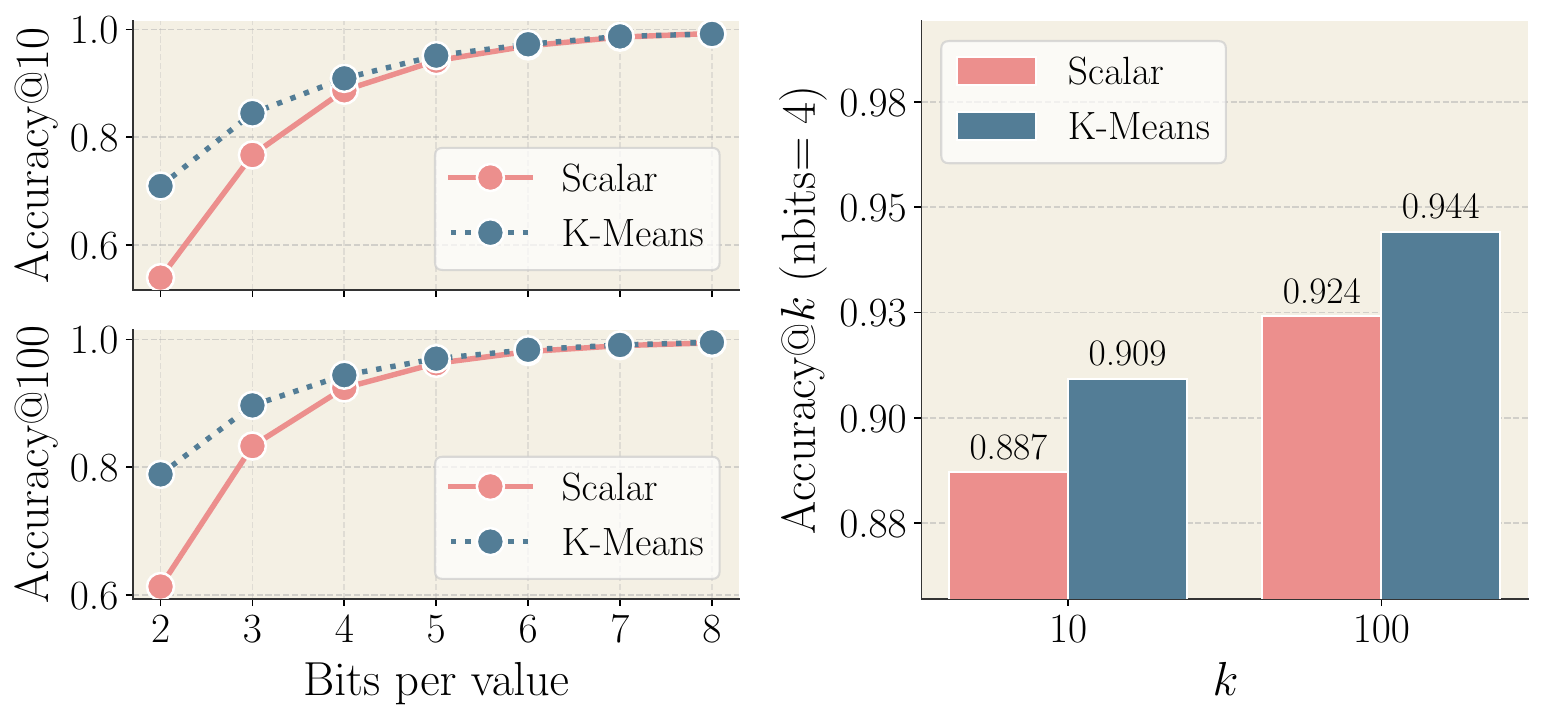}
    \caption{Scalar and $k$-means quantization on \splade-\msmarco.
    Left: Accuracy@$k$ versus bits per value for $k \in \{10, 100\}$
    without oversampling. Right: zoomed view at $4$-bit quantization.
    Four bits recovers most of the $8$-bit accuracy at half the memory. Results obtained on $1000$ sampled queries.}
    \label{fig:scalar_quantization}
\end{figure}

Figure~\ref{fig:scalar_quantization} reports Accuracy@10 for scalar and
$k$-means quantization on the \splade encoding of \msmarco. Here,
Accuracy@10 measures the fraction of the exact top-10 nearest neighbors
that are preserved after quantization; we use this name to distinguish it from qrel-based Recall@10 of \msmarco. Four-bit quantization is the most attractive operating point: it halves the memory footprint of the $8$-bit configuration at only a marginal accuracy cost.
Moreover, $4$-bit quantization admits a particularly efficient SIMD implementation.
The number of codes we can decode is bottlenecked by the registers' ``widening'' capability: once codes are correctly framed into a byte, they need to be widened to \texttt{f32}. Assuming to employ up to AVX256 registers, we can fit up to $8$ values. 

\smallskip
\noindent \textbf{Non-Uniform Quantization}.
Non-Uniform quantization learns a codebook $B_j = {b_j^0, b_j^1, \ldots, b_j^n} $ with $n=2^{\text{nbits}}$ for each component $j$. A generic value $v_j$ is approximated as $ v_j \approx \bar{v}_j = b_j^*$
where $b_j^* = \argmin_i || b_j^i - v_j ||_2 $. For each component, we learn the codebook using a greedy version of univariate k-means, initialized with uniform quantization, and iterate $10$ times. 
Similarly to what we did for scalar quantization, we report the effectiveness of the non-uniform quantization for different numbers of bits, for $k=10, 100$. K-Means outperforms scalar quantization, especially on a low bit budget. We explicit the $4$ bit case in Figure. As shown, k-Means improves by $2$ point the effectiveness with $k=10$ thanks to its more precise representation of the centroids.

We can exploit the same routine to efficiently extract the $4$ bits quantized codes that we presented for scalar quantization. Given the reduced number of bits that we are considering, i.e., $4$, we can use the densify approach for dot product computation. In this case, for each component $j$, we allocate an array of size $|C| \cdot 2^{\text{nbits}}$ that we name $d_q$ (dense query). $d_q$ is populated by precomputing the dot product between the $j$-th component of the query $q_j$ and the codebooks of the $j$-th component. Hence, $d_q^{j + i} = q_j \cdot b_j^i$ $\forall$ $i = 0,.., 2^{\text{nbits}} - 1$, $\forall$ $j=0, .., |C| - 1$. The dot product consists of accumulating the contributions from $d_q$ given the codebook of each document component.
Implementation-wise, this is done with a gather instruction on the $d_q$ array, by computing the access positions as $j * 2^{nbits} + c$, where $j$ is the index of the component and $c$ is the codebook index. These operations can be easily vectorized. The bottleneck of this implementation is the size of the $d_q$ buffer, that is $2^{\text{nbits}}$ times bigger than the densification array used for scalar quantization, which slows down the dot product computation by a $30\%$ in a full-scan scenario.

%% file: sections/04-experiments.tex

\section{Experimental Evaluation}
\label{sec:experiments}


Our experimental evaluation is conducted on \msmarco, a public English passage-retrieval benchmark comprising $8.8$ million passages and $6{,}980$ queries from the \texttt{dev.small} set.
We consider three state-of-the-art learned sparse encoders.

\begin{itemize}[leftmargin=*]
\item \splade~\cite{formal2022splade}.
Each non-zero entry corresponds to the importance weight of a term in the BERT~\cite{devlin2019bert} WordPiece~\cite{wordpiece} vocabulary, which contains $\sim$30{,}000 terms.
\splade achieves MRR@10 $=38.3$ on the \msmarco dev set.
On \msmarco, documents and queries contain, on average, $119$ and $43$ non-zero entries, respectively.

\item \infless~\cite{nardini2025effective}.
\infless belongs to the family of inference-free learned sparse retrievers.
In these models, query processing avoids running a neural encoder at inference time, making query representations inexpensive to compute.
\infless uses a \splade-like encoder for documents, while query-token scores are obtained from a learned lookup table.
We use the \texttt{big} model~\cite{nardini2025effective}, which achieves MRR@10 $=38.8$ on \msmarco.
The resulting representations contain, on average, $387$ non-zero entries per document and $6$ per query.

\item \spladeT~\cite{lassance2024spladev3}.
\spladeT is a more recent and stronger member of the \splade family.
It shares the same vocabulary of \splade, but is trained with improved distillation, hard-negative mining, and regularization, achieving MRR@10 $=40.2$ on the \msmarco dev set.
Documents and queries contain, on average, $170$ and $25$ non-zero entries, respectively.
\end{itemize}

\noindent \textbf{Evaluation Metrics}.
We evaluate indexes along three dimensions: query efficiency, search approximation quality, and memory consumption.
Query efficiency is measured as average end-to-end retrieval latency in microseconds ($\mu$s), excluding query encoding time.
Approximation quality is measured with Accuracy@$k$, for $k \in \{10,100\}$.
Given the exact top-$k$ results computed by exhaustive search, Accuracy@$k$ is the fraction of those results returned by the approximate index.
For each target Accuracy@$k$, we report the fastest configuration among all tested hyperparameter combinations that achieves the target under a given memory budget ($MB$), which is reported as a multiple of the dataset size, i.e., the space required to store the sparse vectors in the forward index by using 16-bit floating-point for values and 16-bit integers for component identifiers. All three encoders we consider build on the BERT vocabulary of $\sim$30{,}000 terms, which implies $|V| \leq 2^{16}$ and allows them to be represented with 16-bit integers.

\vspace{1mm}
\noindent \textbf{Hardware Platform}.
All experiments are run on a server equipped with an Intel Core i9-9940X CPU clocked at $3.30$ GHz ($14$ physical cores, $28$ hardware threads) and 128 GiB of RAM.
Query processing is single-threaded.

\vspace{1mm}
\noindent \textbf{Implementation}.
Our code is written in Rust 1.96.0-nightly. 
The forward index compression methods are integrated in \href{https://github.com/TusKANNy/vectorium}{\texttt{vectorium}}, allowing for a seamless integration into both \href{https://github.com/TusKANNy/seismic}{\seismic} and \href{https://github.com/TusKANNy/kannolo}{\kannolo} codebases.
For \sindi, we use the code released with the original paper,\footnote{https://github.com/Roxanne0321/vsag} as, when privately contacting the authors, they suggested to us that this is faster than the SINDI version integrated in the \texttt{vsag} library.


\subsection{Preliminary Ablation Study}
We now present a detailed comparison performed in isolation of the contributions introduced in Section \ref{sec:methodology}.

\smallskip
\noindent \textbf{Medoids vs. Summary}.
The choice of block representative affects both inverted-index space and candidate-generation effectiveness. Table~\ref{tab:medoids_vs_summary_compact} compares medoids with the original \seismic summaries on \splade at target Accuracy@$10$ levels of $90\%$ and $95\%$. To isolate the effect of the block representative, both strategies use the same forward-index representation, with \dotpacking-encoded components and values quantized using 8-bit per-component scalar quantization ($2.13$ GiB).
\begin{table}[!tb]
    \centering
    \caption{Summary vs. Medoids on \splade.\label{tab:medoids_vs_summary_compact}}
    \small
    \setlength{\tabcolsep}{5pt}
    \begin{tabular}{llrrr}
        \toprule
        \textbf{Target} & \textbf{Strategy} & \textbf{Inv. Idx (GiB)} & \textbf{Latency ($\mu s$)} \\
        \midrule
        \multirow{2}{*}{90\%} & Summary & 1.87 & 162 \\
                              & Medoids & 0.47 & 438 \\
        \midrule
        \multirow{2}{*}{95\%} & Summary & 1.97 & 332 \\
                              & Medoids & 0.67 & 794 \\
        \bottomrule
    \end{tabular}
\end{table}

The results confirm that the medoids strategy achieves a substantial reduction in the memory footprint of the inverted index compared to the summary strategy.
On the other hand, query latency is higher for medoids. This is expected, as summaries provide a tighter upper bound on the score, enabling more aggressive block pruning. However, medoids are preferred in memory-constrained scenarios where reducing the RAM overhead of the inverted index is more critical than absolute retrieval speed.
We complement our analysis by studying the approximation cababilities of summaries and medoids. For every visited block, we compute the ratio of the pruning score (query $\cdot$ summary or query $\cdot$ medoid) to the true block maximum at Accuracy@10 = 95\% (k=10). We first note that the summary (in its alpha-mass pruned form) is already non-conservative: it underestimates the true block maximum in 68.7\% of visited blocks, with a mean ratio of 0.92. Replacing summaries with medoids under the same build and query configuration lowers the mean ratio to 0.86, and Accuracy@10 drops from 95\% to 90\%. The medoid is a more aggressive, but not qualitatively different, approximation of the block bound.
To recover 95\% accuracy with medoids, we relax the heap-factor from 0.9 to 0.7, obtaining the 794 $\mu$s / 2.8 GiB configuration in Table VI (\splade, $MB < 1$, \seismic [new]). At this operating point, the engine scores $4.3\times$ more documents per query ($2$,$875$ vs $664$), which accounts for the higher latency.

\smallskip
\noindent \textbf{Forward Index Compression}.
We now isolate the effect of forward-index component encoding.
Table~\ref{tab:fwd_index_encodings} compares \dotvbyte~\cite{bruch2026forward} with alternative forward-index component encodings (Section~\ref{subsec:compcompr}) within the \seismic pipeline. For each accuracy target, all methods use the same inverted index, pruning parameters, and candidate set, as well as identical vocabulary reordering and 8-bit fixed-point value quantization.
Only component encoding varies.
Therefore, differences in latency are due only to the cost of decoding the forward index and computing exact dot products between the query and the candidate documents.

\begin{table}[]
    \caption{Space/Latency trade-off of different forward index encodings on \splade.\label{tab:fwd_index_encodings}}
    \centering
    \setlength{\tabcolsep}{4pt}
    \newcommand{\spacegain}[1]{{\ \scriptsize\textcolor{gray}{[#1]}}}
    \adjustbox{max width=\columnwidth}{%
    \begin{tabular}{l r@{\hspace{.35\tabcolsep}}l r rrr}
        \toprule
        & \multicolumn{3}{c}{\textbf{Forward Index}}
        & \multicolumn{3}{c}{\textbf{Latency ($\mu$s) }} \\
        \cmidrule(lr){2-4} \cmidrule(lr){5-7}
        \textbf{Method}
        & GiB & & BPE
        & 90\% & 95\% & 99\% \\
        \midrule
        \uncompressed
            & 3.03 & 
            & 24.53 & 179 & 281  & 2$,$080  \\
        \dotvbyte
            & 2.35 & \spacegain{-22.3\%}
            & 19.07 & 154 & 250  & 1$,$695  \\
        \midrule
        \cexpgolomb
            & 1.92 & \spacegain{-36.6\%}
            & 15.55 & 663 & 1$,$207 & 10$,$920 \\
        \cdotpackingdp
            & 2.01 & \spacegain{-33.6\%}
            & 16.29 & 312 & 568  & 5$,$032  \\
        \expgolomb
            & 2.03 & \spacegain{-33.1\%}
            & 16.42 & 734 & 1$,$280 & 11$,$757 \\
        \cdotpacking
            & 2.07 & \spacegain{-31.6\%}
            & 16.78 & 210 & 381  & 3$,$269  \\
        \dotpackingdp
            & 2.07 & \spacegain{-31.6\%}
            & 16.78 & 258 & 463  & 4$,$121  \\
        \dotpacking
            & 2.13 & \spacegain{-29.6\%}
            & 17.28 & 145 & 254  & 1$,$674  \\
        \bottomrule
    \end{tabular}}
\end{table}

The first group of columns reports the space usage of the forward index.
The column ``GiB'' is the total memory occupied by the encoded forward index.
In squared brackets, we report the relative space variation with respect to the uncompressed \uncompressed representation.
The column ``BPE'' reports the average number of bits per vector entry, including both the compressed component identifier and the quantized value.
The second group of columns reports query latency.
For each target Accuracy@$10$ in $\{90\%,95\%,99\%\}$, we report the average per-query latency, in microseconds, obtained when the corresponding forward index encoding is used inside the \seismic pipeline.

The results show that \dotpacking provides the best speed--space trade-off among the tested encodings.
Compared with the uncompressed \uncompressed representation, it reduces the forward index from $3.03$ to $2.13$ GiB, corresponding to a $29.6\%$ space saving, while also improving latency at all accuracy targets.
It is slightly faster than \dotvbyte at $90\%$ and $99\%$ accuracy levels ($145$ vs. $154~\mu$s and $1{,}674$ vs. $1{,}695~\mu$s), and essentially tied at $95\%$ ($254$ vs. $250~\mu$s), while using $9.4\%$ less space.
The more aggressively compressed variants further reduce the index size, but this comes at a substantial latency cost.
For instance, \dotpackingdp reaches $2.07$ GiB and $16.78$ BPE, but is slower than \dotpacking at all targets. By exploiting redundancies across embeddings via reference lists, the clustered-reference variants \cdotpacking and \cdotpackingdp achieve some of the lowest space footprints (down to $2.01$ GiB). While they remain slower than \dotpacking, the double-gather strategy keeps their latency reasonable.
The same trend is even more pronounced for \expgolomb and \cexpgolomb: although they obtain the smallest forward indexes, their scalar decoding makes exact reranking much slower.
This confirms that compression ratio alone is not sufficient for learned sparse retrieval: the encoding must support efficient SIMD-friendly dot-product evaluation.
Overall, \dotpacking is the only method that improves over \dotvbyte in space while preserving, and in some cases slightly improving, query latency. This evidence supports using \dotpacking as the method for compressing components in the forward index in the next sections.

\smallskip
\noindent \textbf{Dot Product Computation}.
We now evaluate the dot product computation methods introduced in Section~\ref{subsec:dotprod}.
The densified kernel remains the fastest option for \splade and \spladeT, where both queries and documents are expanded.
However, this strategy becomes less effective for \infless, where document encodings contain hundreds of non-zero entries, whereas query encodings contain only the tokens appearing in the input text, e.g., $6$ on average on \msmarco. 
In this scenario, the densified kernel still pays the cost of scanning large document supports, even though only a few query terms can contribute to the score.
Therefore, the following analysis compares the densified kernel, the merge kernel, and our \jumpdot on \infless.

\begin{table}[]
\centering
\caption{Effect of the dot product kernel on \infless for $k=10$. Latency ($\mu$s/query) vs memory (GiB) at a target accuracy level.\label{tab:dotprodcomp}
}
\setlength{\tabcolsep}{4pt}
\newcommand{\speedup}[1]{{\! \scriptsize\textcolor{gray}{\,[#1]}}}
\adjustbox{max width=\columnwidth}{%
\begin{tabular}{l
                r@{}l r
                r@{}l r
                r@{}l r}
\toprule
\multirow{2}{*}{\textbf{Method}}
& \multicolumn{3}{c}{\textbf{90\%}}
& \multicolumn{3}{c}{\textbf{95\%}}
& \multicolumn{3}{c}{\textbf{99\%}} \\
\cmidrule(lr){2-4}
\cmidrule(lr){5-7}
\cmidrule(lr){8-10}
& $\mu$s & & GiB
& $\mu$s & & GiB
& $\mu$s & & GiB \\
\midrule
\textsc{Densified}
    & 858 & & 15.7
    & 1{,}253 & & 16.6
    & 3{,}068 & & 23.1 \\
    \textsc{Merge}
      & 628 & \speedup{1.4$\times$} & 15.7
      & 931 & \speedup{1.3$\times$} & 16.6
      & 2{,}307 & \speedup{1.3$\times$} & 23.1 \\
\jumpdot
    & 407 & \speedup{2.1$\times$} & 16.0
    & 594 & \speedup{2.1$\times$} & 16.9
    & 1{,}475 & \speedup{2.1$\times$} & 23.4 \\
\midrule
\dotpacking
    & 2{,}276 & & 6.6
    & 4{,}703 & & 7.2
    & -- & & -- \\
\dotpackingblocked
    & 982 & \speedup{2.3$\times$} & 7.5
    & 1{,}986 & \speedup{2.4$\times$} & 8.0
    & -- & & -- \\
\bottomrule
\end{tabular}}
\end{table}

Table~\ref{tab:dotprodcomp} shows that \jumpdot is consistently faster.
With the uncompressed forward index, it gives a stable $2.1\times$ speedup at all accuracy targets, with less than $2\%$ memory overhead.
As \jumpdot is much faster \textsc{Merge}, we do not implement a \dotpacking-compatible version of \textsc{Merge}.
The gain of \dotpackingblocked is even larger on \dotpacking: $2.3\times$ at $90\%$ and $2.4\times$ at $95\%$ Accuracy@10, at the cost of an $11$--$14\%$ memory overhead caused by the need to store the last component of each block in an uncompressed format.
Overall, the blocked dot product is preferable for inference-free retrievers, where queries are much sparser than documents. For expanded-query encoders such as \splade and \spladeT, we experimentally verify that the densified kernel remains faster, so we use the blocked variant only for \infless in the main comparison.

\begin{figure}[b]
    \centering
    \includegraphics[width=1.0\linewidth]{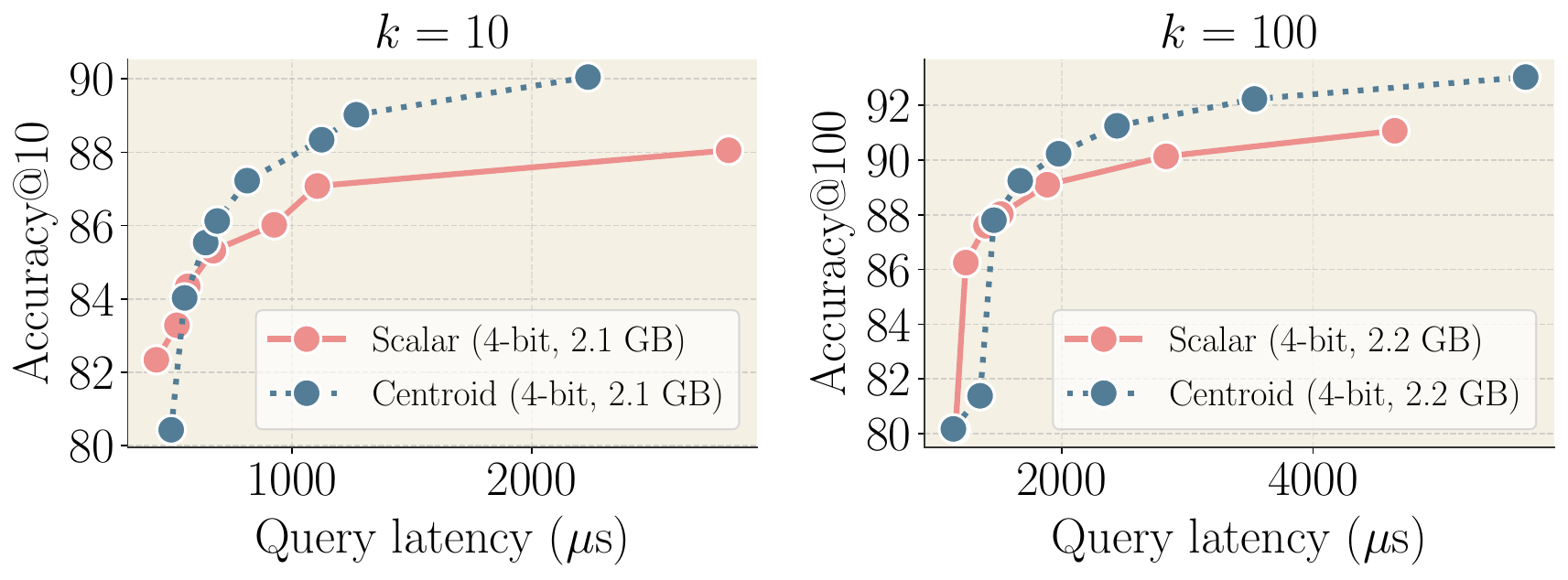}
    \caption{Latency--accuracy trade-off of $4$-bit value quantization on \splade.
    We compare scalar and centroid-based quantization for $k=10$ and $k=100$.
    Both methods use the same compressed component representation and differ only in how values are quantized.\label{fig:quantization_pareto}}
\end{figure}

\subsection{Experimental Results}
\label{subsec:sota}
We experimentally evaluate the following variants of \seismic and \kannolo. For \seismic, we include: 1) the original \seismic with summaries and an uncompressed \texttt{f16} forward index, and 2) \seismic [new], which differs in i) how block representatives are chosen, with summaries or with medoids, ii) the compression applied to the forward index, namely no compression (\texttt{f16}) or \dotpacking, and iii) the dot product used. For \kannolo, we include: 1) the original \kannolo exploiting edge compression using StreamVByte, and 2) \kannolo [new], which also exploits \dotpacking for forward index compression.
We compare our solutions with \sindi~\cite{li2025sindi}, \seismic~\cite{bruch2024seismic,Bruch2024SeismicWave}, and a sparse version of \hnsw implemented in the \kannolo~\cite{kannolo} library, which is much faster than the two winners of the BigANN Challenge@NeurIPS23, i.e., \textsc{PyAnn} and \textsc{sHnsw}~\cite{kannolo}. We also tested \supblock, an improved version of \textsc{Bmp}. \supblock fails to achieve accuracy values $> 98\%$ due to the query and document quantization applied during query processing. As its performance at lower recall levels always lags behind \sindi (around $2\times$ slower), we do not report it in our evaluation. We also include \qdrant~\cite{qdrant}, a widely adopted production vector-search engine for exact search, configured with \texttt{u8} scalar quantization to minimize its memory footprint.

We report the results on \splade, \infless, and \spladeT for $k \in \{10,100\}$.
For each encoder and value of $k$, we group configurations by memory budget, expressed as a multiple of the dataset size. We consider two scenarios i) memory budget $\geq 1$ and memory budget $< 1$.
\sindi stores component identifiers using $32$-bit integers and values using floating-point numbers; therefore, its memory consumption is always above $2\times$ the dataset size in our setting. Our methods applying forward index compression are the only ones that can achieve a memory budget $< 1$, together with \qdrant. 
For each method, we perform a grid search on building and search parameters. We report the fastest configuration that achieves each accuracy target ($90\%$, $95\%$, $99\%$) while respecting the memory budget.

\smallskip
\noindent \textbf{Hyperparameters}. For \sindi, we fix the window size to $60{,}000$ and enable index reordering. We explore the document mass-retention ratio $\alpha \in [0.2, 1.0]$ step $0.2$ at building time; at search time, we use the query term retention ratio $\beta \in [0.2, 1.0]$ step $0.2$ and the candidate pool size $\gamma \in \{50, 100, 150, 200, 300, 500, 700, 1000, 1500\}$ for $k{=}10$, $\{100, 150, 200, 300, 500, 700, 1000, 1500, 2000\}$ for $k{=}100$. Note that $\alpha{=}1.0$ and $\beta{=}1.0$ disable document and query pruning, respectively. These parameters were derived by extending the original set provided by the authors~\cite{li2025sindi}. For \kannolo, we use $M \in \{8, 16, 32\}$  and \texttt{ef-construction}${=}2000$; at search time, we used \texttt{ef-search} $\in$ $[10, 16]$ step 1 for k=10, $[100, 160]$ step $10$ for k=100, $\lambda \in [0.01, 0.13] $ step $0.01$, where $\lambda$ is the early termination parameter. In \kannolo, edges are compressed using streamvbyte to yield a fair comparison with the other methods~\cite{benocci2026lossless}. For \seismic,  at building time we explore the posting list length $n_p \in \{1000, 2000, 3000, 4000, 5000\}$ (extended up to $12{,}000$ on \infless), the centroid fraction $\in \{0.05, 0.1, 0.2\}$, and the maximum list-length factor $\in \{2, 3, 4, 6\}$. For the energy-preserving summaries of \seismic, we further explore the summary energy $\eta \in \{0.4, 0.5, 0.6\}$; \seismic replaces the quantized summaries with cluster medoids, which require no additional parameter. At search time, we use the query cut $q_c \in \{2, \dots, 12\}$ and the heap factor $h \in [0.7, 1.0]$ step $0.1$ for the summary-based variants, $h \in [0.4, 0.8]$ step $0.05$ for the medoid-based ones.

\begin{table*}[]
    \centering
    \caption{Per-query latency ($\mu$s) and memory footprint (GiB) at $90$/$95$/$99$\% accuracy.
    Methods in bold are proposed in this work.
    \label{table:results1}}
    \setlength{\tabcolsep}{3.5pt}
    \adjustbox{max width=\textwidth}{%
    \begin{tabular}{l l rr rr rr @{\hspace{1.4em}} rr rr rr}
        \toprule
        \multirow{4}{*}{\textbf{Mem. Budget}} & \multirow{4}{*}{\textbf{Method}}
          & \multicolumn{6}{c}{$\bm{k=10}$} & \multicolumn{6}{c}{$\bm{k=100}$} \\
        \cmidrule(lr){3-8} \cmidrule(lr){9-14}
        & & \multicolumn{2}{c}{90\%} & \multicolumn{2}{c}{95\%} & \multicolumn{2}{c}{99\%}
            & \multicolumn{2}{c}{90\%} & \multicolumn{2}{c}{95\%} & \multicolumn{2}{c}{99\%} \\
        \cmidrule(lr){3-4}\cmidrule(lr){5-6}\cmidrule(lr){7-8}\cmidrule(lr){9-10}\cmidrule(lr){11-12}\cmidrule(lr){13-14}
        & & $\mu$s & GiB & $\mu$s & GiB & $\mu$s & GiB & $\mu$s & GiB & $\mu$s & GiB & $\mu$s & GiB \\
        \midrule
        \multicolumn{14}{c}{\textit{\splade on \msmarco} --- Memory Budget $1.0$ =  4.0 GiB} \\
        \midrule
        \multirow{4}{*}{$\geq 1.0$}
          & \sindi               & 773 & 9.7 & 1{,}131 & 9.7 & 2{,}204 & 10.2 & 1{,}968 & 10.2 & 2{,}746 & 10.7 & 4{,}604 & 10.7 \\
          & \seismic          & 197 & 4.1 & 275  & 5.8 & 621  & 7.8  & 739  & 7.3  & 1{,}129 & 8.4  & 2{,}862 & 8.4  \\
          & \kannolo    & 299 & 4.6 & 431 & 4.6 & 999 & 4.9 & 823     & 4.6 & 942     & 4.6 & 2{,}512 & 4.9  \\

        & \textbf{\seismic [new]}  & 162 & 4.0 & 332 & 4.1 &640& 6.8 & 764  & 4.2 & 1{,}849 & 4.2 & 2{,}943 & 7.4  \\
        \midrule
        \multirow{3}{*}{$ < 1.0$}
                & \qdrant  & &  &  &  & 56{,}970 & 2.2 &  & &  & & 56{,}690 & 2.2 \\
        
        & \textbf{\seismic [new]} & 438 & 2.6 & 794 & 2.8 & 2{,}337 & 3.8 & 1{,}178 & 2.9 & 2{,}116 & 3.0 & --   & --  \\
             & \textbf{\kannolo [new]} & 301 & 2.7 & 434 & 2.7 & --  & --  & 776 & 2.7 & 900 & 2.7 & 4$,$350 & 3.1    \\
        \midrule
        \midrule
        \multicolumn{14}{c}{\textit{\infless on \msmarco} --- Memory Budget $1.0$ =  12.8 GiB} \\
        \midrule
        \multirow{5}{*}{$ \geq 1.0$}
          & \sindi                         & 1{,}909 & 29.0 & 3{,}673 & 31.3 & 8{,}926 & 31.3 & 5{,}821 & 30.0 & 9{,}545 & 30.0 & 23{,}843 & 31.0 \\
          & \seismic                     & 858  & 15.7 & 1{,}253 & 16.6 & 3{,}068 & 23.1 & 2{,}996 & 20.1 & 3{,}284 & 20.1 & 9{,}258  & 26.1 \\
          & \kannolo & 5{,}779 & 13.3 & 27{,}642 & 13.3 & -- & -- & 8{,}069 & 13.3 & 43{,}173 & 13.3 & -- & -- \\                              
          
          & \textbf{\seismic [new]} & 463  & 8.9  & 690  & 10.9  & 1{,}475 & 23.4 &  1{,}447 & 13.8 & 1{,}657 & 13.8 & 8{,}792 & 21.6 \\
        & \textbf{\kannolo [new]}   & 3{,}312 & 13.5 & 17{,}161 & 13.5 & -- & -- & 4{,}500 & 13.5 & 26{,}623 & 13.5 & -- & -- \\

        \midrule
          \multirow{3}{*}{$< 1.0$}& 
       \qdrant & & &  & & 33{,}506 & 6.4 &  &  &  &  & 33{,}411 & 6.4 \\
          
      &    \textbf{\seismic [new]} & 982 & 7.5 & 1{,}986 & 8.0 & 11{,}832 & 11.2 & 2{,}206 & 7.8 & 4{,}238 & 8.0 & - & -\\
  & \textbf{\kannolo [new]} & 3{,}616 &  6.9 & 39{,}749 &  6.9 & -- & -- & 5{,}175 &  6.9 & 25{,}110 &  6.9 & -- & -- \\
       
        \midrule
        \midrule
        \multicolumn{14}{c}{\textit{\spladeT on \msmarco} --- Memory Budget $1.0$ = 6.0 GiB} \\
        \midrule
        \multirow{4}{*}{$\geq 1.0$}
          & \sindi              & 931 & 13.9 & 1{,}494 & 13.9 & 2{,}150 & 13.9 & 2{,}156 & 13.9 & 3{,}112 & 13.9 & 5{,}357 & 15.4 \\
          & \seismic            & 158 & 8.6  & 199  & 9.0  & 462  & 11.1 & 981  & 9.2  & 1{,}281 & 10.6 & 3{,}114 & 11.1 \\
          
            & \kannolo & 445 & 6.4 & 636 & 6.4 & 2{,}327 & 6.4 & 1{,}425 & 6.4 & 1{,}657 & 6.4 & 7{,}729 & 6.4 \\
          & \textbf{\seismic [new]} & 162 & 5.6  & 211  & 6.9  & 949  & 9.1  & 993  & 6.3  & 1{,}289 & 6.6  & 3{,}679 & 9.4  \\


        \midrule
           \multirow{3}{*}{$< 1.0$}& \qdrant  & & & & & 44{,}182 & 3.1 & & & &  & 44{,}280 & 3.1 \\

           & \textbf{\seismic [new]} & 594 & 3.7 & 978 & 3.7 & 3{,}324 & 5.4 & 1{,}517 & 3.7 & 2{,}441 & 4.1 & 12{,}676 &  5.4  \\
        & \textbf{\kannolo [new]} & 417 & 3.4 & 782 & 3.4 & -- & -- & 924 & 3.4 & 1{,}402 & 3.4 & 20{,}531 & 3.7 \\
          
        \bottomrule
    \end{tabular}}
\end{table*}

\vspace{1mm}
\noindent\textbf{Results}.
Table~\ref{table:results1} reports the results of our evaluation. Among the uncompressed baselines, \seismic is always the fastest one, being from $1.6\times$ to $7.5\times$ faster than \sindi. \kannolo is also consistently faster than \sindi, excluding the \infless dataset, where \kannolo is the slowest solution. As previously reported in the literature~\cite{bruch2024seismic}, graph-based data structures struggle with heavily unbalanced sparsity patterns between queries and documents. 
On \splade, \seismic [new] (MB $\geq 1)$ yields almost the same efficiency of \seismic, saving up to $4.2$ GiB. \seismic [new] can be compressed more aggressively (MB $< 1)$ by replacing summaries with medoids. This allows saving up to $5.4$ GiB at the cost of a slowdown of around $1.6$--$3.8\times$. \kannolo [new] is as fast as its base version, but it is $41\%$ more memory-efficient, reducing space usage from $4.6$ to $2.7$ GiB.

On \spladeT, we observe the same speed--space trade-off.
In the $MB \geq 1$ regime, \seismic [new] is close to the latency of the original \seismic, while consistently reducing the memory footprint.
The gap becomes more visible as the accuracy target increases: the original \seismic keeps an advantage in raw speed, but \seismic [new] achieves comparable accuracy with a substantially smaller index.
When moving below the dataset-size budget ($MB < 1$), the uncompressed \seismic configuration is no longer applicable, whereas \seismic [new] continues to reach the $90\%$ and $95\%$ targets for both values of $k$.
This shows that the proposed compression and medoid-based selection move \seismic to a lower-memory regime while preserving practical query efficiency.
A similar trend holds for \kannolo [new], which retains the latency profile of the original graph-based index at moderate accuracy while nearly halving its memory footprint.

On \infless, \seismic [new] benefits from the highly asymmetric sparsity pattern of inference-free retrieval.
Thanks to \jumpdot and \dotpackingblocked kernels, it is both the fastest and the most memory-efficient specialized method across the reported targets. 
Conversely, \kannolo and \kannolo [new] are much less competitive, confirming that graph-based traversal is less robust when queries are extremely sparse, and documents remain highly expanded, and are both slower than \sindi, even when equipped with the highly efficient \dotpackingblocked kernel. 

Finally, \qdrant confirms that generic exact sparse-vector search can be made compact, but its latency is two orders of magnitude above specialized approximate LSR indexes.

\subsection{Ultra Compact Index}
\label{subsec:ultra_compact}

We study how far the forward index can be compressed by combining component compression with $4$-bit value quantization. 
This experiment revisits the memory bottleneck highlighted in the introduction: the original \seismic index on \splade reaches microsecond-level latency, but requires about $7$ GiB of memory to operate at high accuracy. 
Here, we push the total index footprint (forward plus inverted index) close to $2$ GiB, roughly half of the dataset-size budget, while still preserving about $90\%$ Accuracy@$10$.

We compare the two value-quantization strategies described in Section~\ref{subsec:valcompr}. 
Scalar quantization assigns each value to one of $16$ uniformly spaced bins for its component, whereas centroid-based quantization learns $16$ reconstruction values independently for each component. 
Both variants use the same compressed component representation and the same \seismic retrieval pipeline.
Figure~\ref{fig:quantization_pareto} reports the latency--accuracy trade-off for $k=10$ and $k=100$. 
Centroid quantization consistently improves the Pareto frontier: for $k=10$, it reaches about $90\%$ Accuracy@$10$ with a total index footprint of $2.1$ GiB, while for $k=100$ it exceeds $92\%$ Accuracy@$100$ using $2.2$ GiB.
Scalar quantization is slightly faster in some low-latency configurations, but saturates earlier and does not reach the same high-accuracy regime.
Overall, $4$-bit value quantization enables an ultra-compact index with a total index footprint close to $2$ GiB, corresponding to a memory budget of about $0.5$. 
Although this configuration is not meant to dominate the fastest high-memory operating points, it shows that learned sparse retrieval can remain practically effective in a memory regime far below that of the original \seismic design.

%% file: sections/05-conclusions.tex

\section{Conclusions and Future Work}
\label{sec:conclusions}
We studied how to substantially reduce the memory footprint of learned sparse retrieval indexes without sacrificing efficiency. Building on \seismic, we revisited both components of its two-level design. On the selector side, we replaced the costly per-block summaries with \emph{medoids}, collapsing blocks from sparse vectors into a single document identifier. On the forward index side, we proposed a compression scheme that acts on both components, encoded with \dotpacking---a SIMD-fused bit-packing of reordered $\Delta$-gaps---and values, quantized with compact per-component $4$-bit codebooks. Our forward index contribution is orthogonal to \seismic and can be exploited by any system that relies on forward-index-based candidate scoring. Our evaluation on \msmarco with three learned sparse encoders confirms that these techniques significantly improve the speed-space trade-off:
\seismic is up to $5.3\times$ faster than \sindi{} while using about $3\times$ less memory. Moreover, in the most memory-constrained regime, it remains up to $1.9\times$ faster while using up to $3.9\times$ less memory.
As future work, we intend to study whether multiple, or learned, block representatives can recover the tighter pruning signal of summaries while retaining their compactness.